\documentclass[12pt]{article}
\usepackage{epsfig}
\usepackage{graphicx}
\usepackage{a4}
\usepackage{latexsym}
\usepackage{cite}
\usepackage{slashed}
\usepackage{color}
\usepackage{colordvi,scalefnt}

\usepackage{pslatex}
\usepackage[latin1]{inputenc}
\usepackage[T1]{fontenc}

\newcommand{\as}{\alpha_{\rm s}}

\def\MSbar{\overline{\mathrm{MS}}}
\def\ep{\epsilon}
\def\z#1{{\zeta_{#1}}}
\def\ca{{C^{}_A}}
\def\cf{{C^{}_F}}
\def\tf{{T^{}_F}}
\def\nf{{n^{}_{\! f}}}

\def\lm{\mathrm{L}_m}
\def\ls{\mathrm{L}_s}

\def\ly{\mathrm{L}_y}
\def\s#1#2{\mathrm{S}_{#1,#2}(x)}
\def\li#1{\mathrm{Li}_{#1}(x)}

\newcommand{\brk}{\right. \nonumber \\ && \left.}
\newcommand{\ibrk}{\right. \right. \nonumber \\ && \left. \left.}
\newcommand{\dibrk}{\right. \right. \right. \nonumber \\ && \left. \left. \left.}

\newcommand{\text}{\textstyle}
\newcommand{\bea}{\begin{eqnarray}}
\newcommand{\eea}{\end{eqnarray}}

\def\Ione#1{\mbox{\boldmath I}_{#1}^{(1)}(\ep)}

\def\nnlo{{\rm NNLO}}

\def\CF{C_F}

\def\A{{\cal A}}

\begin{document}
\setlength{\parskip}{0.2cm} \setlength{\baselineskip}{0.55cm}

\begin{titlepage}
\vspace{1.5cm}
\begin{center}
\LARGE {\bf
W Pair Production at the LHC \\[0.5ex]
II. One-loop Squared Corrections in the High Energy Limit
}\\
\vspace{2.2cm}
\large
G. Chachamis$^{a}$, M. Czakon$^{a,b}$ and D. Eiras$^{a}$\\
\vspace{1.4cm}
\normalsize
{\it
$^{a}$Institut f\"ur Theoretische Physik und Astrophysik, Universit\"at W\"urzburg \\[0.5ex]
Am Hubland, D-97074 W\"urzburg, Germany \\[.5cm]
$^{b}$Department of Field Theory and Particle Physics,
Institute of Physics \\[0.5ex]
University of Silesia, Uniwersytecka 4, PL-40007 Katowice,
Poland \\[.5cm]}

\vspace{4cm}

{\large {\bf Abstract}}

\end{center}
\noindent
We present the result for the one-loop squared virtual QCD
corrections to the W boson pair production in the 
quark-anti-quark-annihilation channel in the limit where all 
kinematical invariants are large compared to the mass of the W boson.
The infrared pole structure is in agreement with the
prediction of Catani's general formalism for the singularities of
QCD amplitudes.

\vspace{-0.2cm}

\vspace{2.0cm}
\end{titlepage}

\newpage

\section{Introduction}
\label{sec:intro}

The Large Hadron Collider (LHC) is expected to have a huge impact
on particle physics phenomenology.
Most of the processes which will be studied at the LHC need to be
calculated at least to next-to-leading order (NLO) in QCD whereas there
are some for which a theoretical prediction is needed to 
next-next-to-leading order (NNLO).
Electroweak gauge boson pair production falls into the latter category.
One of the reasons is that the 
increase of the centre-of-mass energy at the LHC 
with respect to the Tevatron 
from 1.96 TeV to 14 TeV will result in a huge boost of the available
data.

The importance of  hadronic W-pair production is two-fold. Firstly, it is a
process which allows the measurement of the vector boson trilinear couplings 
and therefore a comparison with the Standard Model (SM) predictions.
Most attempts to model New Physics, such as 
Supersymmetry and Extra-dimensions in all
variations, should be able to
explain any 
deviations by consistently adjusting the anomalous
couplings and/or by 
incorporating decays of new particles into vector boson 
pairs~\cite{tevatron}. 

Secondly, hadronic W pair 
production is important for investigations of the nature of the
Electroweak symmetry mechanism by contributing the 
dominant background
for the Higgs boson mediated process 
(see Refs.~\cite{Spira:1995rr,Dawson:1990zj,Harlander:2002wh,Anastasiou:2002yz,
Ravindran:2003um,Catani:2001cr,
Davatz:2004zg,Anastasiou:2004xq,Anastasiou:2007mz,
Grazzini:2008tf,Bredenstein:2006rh,kauer1,kauer2}),
\bea
p p \rightarrow H \rightarrow W^* W^* 
\rightarrow l {\bar \nu} {\bar l}' \nu' \, , \nonumber
\eea
in the Higgs mass range between 140 GeV < $M_{H}$ < 180
GeV~\cite{dittmardreiner}. 

The interest in hadronic W pair production is well displayed by
the fact that the Born cross 
section was calculated some thirty years ago~\cite{brown}.
The NLO QCD corrections were computed in the 90's and seen to contribute
a 30\%~\cite{ohn,fri,dixon1,dixon2,campbell}.  
Next,
soft gluon resummation
effects were considered in Ref.~\cite{grazzini:2006}
whereas massless fermion-boson scattering was studied at NNLO
in Ref.~\cite{Anastasiou:2002zn}.
The first step towards a complete NNLO study is the
computation of the NNLO two-loop virtual corrections
in a high energy expansion, $M_{W}^2 \ll$ s, t, u~\cite{Chachamis:2008yb}.
The purpose
of the present paper is to complete this study by providing the
one-loop squared contributions in the same limit.

The method used in \cite{Chachamis:2008yb}, 
and already sketched in
\cite{qqTT,ggTT,Czakon:2004wm,Czakon:2006pa,Actis:2007gi}, 
is somehow different to the one used for the present work. 
In fact this time we have used the helicity matrix formalism
to reduce the problem to a small set of integrals.
We treat these again with Mellin-Barnes
representations~\cite{Smirnov:1999gc,Tausk:1999vh} 
which are constructed by means of the {\tt
MBrepresentation} package~\cite{MBrepresentation} 
and then analytically continued in
the number of space-time dimensions $D = 4 -2 \epsilon$ using the
{\tt MB} package~\cite{Czakon:2005rk}. 
After the asymptotic expansion in the mass
parameter, contours are closed and integrals finally resummed either with
the help of {\tt XSummer}~\cite{Moch:2005uc} or the {\tt PSLQ}
algorithm~\cite{pslq:1992}. 

In Section~\ref{sec:notation} we introduce our notation, in 
Section~\ref{sec:catani} 
we verify the correctness of the infrared pole structure 
by comparing with the Catani prediction~\cite{catani}. We
present our results in Section~\ref{sec:results} after which we conclude
in Section~\ref{sec:conclusions}.

\section{Notation}
\label{sec:notation}

Although the notation adopted here is identical to that of
Ref.~\cite{Chachamis:2008yb}, 
we shall recapitulate it for completeness.
The charged vector-boson 
production in the leading partonic scattering process
corresponds to
\begin{equation}
\label{eq:qqWW}
q_j(p_1) + {\overline q}_j(p_2) 
\:\:\rightarrow\:\: W^-(p_3,m) + W^+(p_4,m) \, ,
\end{equation}
where $p_i$ denote 
the quark and W momenta, $m$ is the mass of the W boson and
j is a flavour index.
We are considering down-type quark scattering in our paper. Obtaining
the corresponding result for up-type quark scattering is actually trivial as
we will show in the following.
Energy-momentum conservation implies
\begin{equation}
\label{eq:engmom}
p_1^\mu+p_2^\mu = p_3^\mu+p_4^\mu \, .
\end{equation}
We consider
the scattering amplitude ${\cal M}$ for the process~(\ref{eq:qqWW})
at fixed values of the external parton momenta $p_i$, thus $p_1^2 =
p_2^2 = 0$ and $p_3^2 = p_4^2 = m^2$.
The amplitude ${\cal M}$ may be written as a 
series expansion in the strong coupling $\as$,
\begin{eqnarray}
  \label{eq:Mexp}
  | {\cal M} \rangle
  & = &
  \biggl[
  | {\cal M}^{(0)} \rangle
  + \biggl( {\as \over 2 \pi} \biggr) | {\cal M}^{(1)} \rangle
  + \biggl( {\as \over 2 \pi} \biggr)^2 | {\cal M}^{(2)} \rangle
  + {\cal O}(\as^3)
  \biggr]
\, ,
\end{eqnarray}
and we define the expansion parameter in powers of $\as(\mu^2)
/ (2\pi)$ with $\mu$ being the renormalisation scale. We work in
conventional dimensional regularisation, $d=4-2 \ep$, in the
$\MSbar$-scheme for the coupling constant renormalisation.

We explicitly relate the bare (unrenormalised) coupling $\as^{\rm{b}}$
to the renormalised coupling $\as$ by
\begin{eqnarray}
\label{eq:alpha-s-renorm}
\as^{\rm{b}} S_\epsilon \: = \: \as
\biggl[
   1
   - {\beta_0 \over \epsilon} \biggl( {\as \over 2 \pi} \biggr)
   + \left(
     {\beta_0^2 \over \epsilon^2}
     - {1 \over 2} {\beta_1 \over \epsilon}
     \right) \biggl( {\as \over 2 \pi} \biggr)^2
  + {\cal O}(\as^3)
  \biggr]
\, ,
\end{eqnarray}
where we set the factor 
$S_\epsilon=(4 \pi)^\ep \exp(-\ep \,\gamma_{\rm E}) = 1$
for simplicity and $\beta$ is the QCD $\beta$-function known at present up to
the four-loop level
\cite{vanRitbergen:1997va,Czakon:2004bu}
\begin{eqnarray}
\label{eq:betafct}
\beta_0 = {11 \over 6}\*\ca - {2 \over 3}\*\tf\*\nf \, ,
\qquad
\beta_1 = {17 \over 6}\*\ca^2 - {5 \over 3}\*\ca\*\tf\*\nf - \cf\*\tf\*\nf \, .
\end{eqnarray}
The color factors in a non-Abelian ${\rm{SU}}(N)$-gauge theory are
$\ca = N$, $\cf = (N^2-1)/2\*N$ and $\tf = 1/2$.
Throughout this paper, $N$ denotes the number of colors and
$\nf$ the total number of flavors.

For convenience, we define the function ${\cal A}(\epsilon, m, s, t, \mu)$
for the squared amplitudes summed over spins and colors as
\begin{eqnarray}
\label{eq:Msqrd}
\overline{\sum |{\cal M}({q_j + {\overline q}_j \to  W^+ + W^-} )|^2}
&=&
{\cal A}(\epsilon, m, s, t, \mu)
\, .
\end{eqnarray}
${\cal A}$ is a function of the Mandelstam variables $s$, $t$ and $u$ given by
\begin{equation}
\label{eq:Mandelstam}
s = (p_1+p_2)^2\, , \qquad
t  = (p_1-p_3)^2 - m^2\, , \qquad
u  = (p_1-p_4)^2 - m^2\, ,
\end{equation}
and has a perturbative expansion similar to Eq.~(\ref{eq:Mexp}),
\begin{equation}
\label{eq:Aexp}
{\cal A}(\epsilon, m, s, t, \mu) = 
\left[
  {\cal A}^{(0)}
  + \biggl( {\as \over 2 \pi} \biggr) {\cal A}^{(1)}
  + \biggl( {\as \over 2 \pi} \biggr)^2 {\cal A}^{(2)}
  + {\cal O}(\as^{3})
\right]
\, .
\end{equation}
In terms of the amplitudes the expansion coefficients in Eq.~(\ref{eq:Aexp})
may be expressed as
\begin{eqnarray}
\label{eq:A4def}
{\cal A}^{(0)} &=&
\langle {\cal M}^{(0)} | {\cal M}^{(0)}\rangle \, , \\
\label{eq:A6def}
{\cal A}^{(1)} &=& \left(
\langle {\cal M}^{(0)} | {\cal M}^{(1)} \rangle + \langle {\cal M}^{(1)} | {\cal M}^{(0)} \rangle
\right)\, , \\
\label{eq:A8def}
{\cal A}^{(2)} &=& \left(
\langle {\cal M}^{(1)} | {\cal M}^{(1)} \rangle
+ \langle {\cal M}^{(0)} | {\cal M}^{(2)} \rangle + \langle {\cal M}^{(2)} | {\cal M}^{(0)} \rangle
\right)\, .
\end{eqnarray}
Note that in the above equations, only the two-loop amplitude needs
to be renormalised.

${\cal A}^{(0)}$ is given by
\begin{eqnarray}
{\cal A}^{(0)} &=& N \left \{ 
c_1 \left[ 16 (1- \epsilon)^2 \frac{x}{\left (1-x \right )} 
+ 4 (3 - 4\epsilon) \frac{1}{m_s} 
+ \frac{4 x\,(1-x)}{m_s^2}\right ] 
\brk
+ c_2 \left [ -24 + 16 x + 16 \epsilon 
\left ( 2 - x \right ) + 4 \frac{(3 - 4 \epsilon) 
- 2 x (1-x)}{m_s} +  \frac{4 x \,(1-x)}{m_s^2}\right ] 
\brk
+ c_3 \left [ - 24 \left ( 1-x\, (1-x)\right ) 
+ 16 \epsilon \left ( 2 - x \, (1-x)\right ) 
+ \frac{6 - 8 \epsilon - 8 x \, (1-x)}{m_s} 
+ \frac{2 x \, (1-x)}{m_s^2} \right ] \right \} \, ,
   \nonumber \\ 
\end{eqnarray}
where we have defined $x = -\frac{t}{s}$, $m_s = \frac{m^2}{s}$ and only the
leading physical powers (i.e. down to the constant) in the $m_s$-expansion are
retained. Notice that, once the actual values of the $c_i$ are substituted, the
terms singular in $m_s$ cancel as required by unitarity.
This will be the case  for the
one-loop squared expression as well. The coefficients $c_1$,
$c_2$ and $c_3$ are in their essence
combinations of EW coupling constants defined as
\begin{eqnarray}
c_1 &=& \frac{g_{WL}^4}{4} \, , \nonumber \\
c_2 &=& \frac{1}{4 \, s_w^2}\left(Q_q+2 g_{ZL}^q \frac{c_w }{s_w \left (
      1-\frac{M_Z^2}{s}\right )}\right )
\, , \nonumber \\
c_3 &=& \frac{c_w^2}{s_w^2 (1-\frac{M_Z^2}{s})^2} \left (  ( g_{ZA}^q)^2 +
  \left (g_{ZV}^q + Q_q \frac{s_w \left ( 1-\frac{M_Z^2}{s} \right )}{c_w} 
\right )^2 \right ) \, .
\end{eqnarray}

The expressions for ${\cal A}^{(1)}$ have been presented e.g. in 
Refs.~\cite{ohn, fri}, whereas 
the real part of the full two-loop contributions, namely
the result for the last two terms in Eq.~(\ref{eq:A8def}) 
in the high energy limit, was presented in Ref.~\cite{Chachamis:2008yb} 
(the leading color coefficient of 
$\langle {\cal M}^{(0)} | {\cal M}^{(2)} \rangle$ was discussed
in Ref.~\cite{Chachamis:2007cy}).
Here we provide the result for the one-loop$\otimes$one-loop 
result in the high energy limit, namely the NNLO
contribution $\langle {\cal M}^{(1)} | {\cal M}^{(1)}
\rangle$ in ${\cal A}^{(2)}$.

In order to compute the $\langle {\cal M}^{(1)} | {\cal M}^{(1)}
\rangle$ we will use the helicity matrix formalism,
namely we will express  the result in terms of helicity
amplitudes, ${\mathcal M}^g(\lambda_1, \lambda_2, s, t)$.
The quark and anti-quark have opposite helicities in the
centre-of-mass system so one helicity label above, $g = \pm 1$,
suffices. $\lambda_1$ and $\lambda_2$ stand for the
helicities of the W$^+$ and W$^-$ respectively.

Starting from the one-loop amplitude,
the initial expression can be
rearranged as 
\bea
\label{eq:hme}
| {\cal M}^{(1)} \rangle = \sum_{i,j,g}  C_i (s,t,u) 
{\mathcal I}_i^j (s,t,u;\mu^2) {\mathcal M_j}(\{p_k\},g)\, , 
\eea
where the $C_i$ are coefficients, the ${\mathcal I}_i^j$ are one-loop
dimensionally regularized scalar
integrals, ${\mathcal M_j}$ are helicity matrix elements, g = $\pm$ and $k =
1,...,4$. 
The ten helicity matrix elements ${\mathcal M_j}(p_k,g) = 
{\mathcal  M_j^g}$ have been taken as defined in Ref.~\cite{Diener:1997nx} 
(see also~\cite{Denner:1988tv}):
\bea
{\mathcal M}_0^g &=& {\overline v }(p_2) \, \slashed{\varepsilon_1} (\slashed{p_3} -
\slashed{p_2}) 
\slashed {\varepsilon_2}
{\mathcal P}_g \, u(p_1)  \, , \nonumber \\
{\mathcal M}_1^g &=& {\overline v}(p_2) \, \slashed{p_3} {\mathcal P}_g \, u(p_1) \,
\varepsilon_1 \cdot \varepsilon_2 \, , \nonumber \\
{\mathcal M}_2^g &=& {\overline v}(p_2)\, \slashed{\varepsilon_1} {\mathcal P}_g \, u(p_1)
\, \varepsilon_2 \cdot p_3 \, , \nonumber \\
{\mathcal M}_3^g &=& -{\overline v}(p_2)\,\slashed{\varepsilon_2} {\mathcal P}_g \, u(p_1)
\, \varepsilon_1 \cdot p_4 \, , \nonumber \\
{\mathcal M}_4^g &=& {\overline v}(p_2)\, \slashed{\varepsilon_1} {\mathcal P}_g \, u(p_1)
\, \varepsilon_2 \cdot p_1\, , \nonumber \\
{\mathcal M}_5^g &=& -{\overline v}(p_2)\, \slashed{\varepsilon_2} {\mathcal P}_g \, u(p_1)
\, \varepsilon_1 \cdot p_2\, ,  \\
{\mathcal M}_6^g &=& {\overline v}(p_2)\, \slashed{p_3} {\mathcal P}_g \, u(p_1) \,
\varepsilon_1 \cdot p_2 \, \varepsilon_2 \cdot p_1
\, , \nonumber \\
{\mathcal M}_7^g &=& {\overline v}(p_2)\, \slashed{p_3}{\mathcal P}_g \, u(p_1) \,
\varepsilon_1 \cdot p_2 \, \varepsilon_2 \cdot p_3 \, , \nonumber \\
{\mathcal M}_8^g &=& {\overline v}(p_2)\, \slashed{p_3}{\mathcal P}_g \, u(p_1)\,
\varepsilon_1 \cdot p_4 \, \varepsilon_2 \cdot p_1 \, , \nonumber \\
{\mathcal M}_9^g &=& {\overline v}(p_2)\, \slashed{p_3} {\mathcal P}_g \, u(p_1)\,
\varepsilon_1 \cdot p_4 \, \varepsilon_2 \cdot p_3 \, , \nonumber
\eea
where ${\mathcal P}_g = {\mathcal P}_{\pm} = \frac{1 \pm \gamma_5}{2}$.
All colour indices as well 
as the arguments of the polarization vectors,
$\varepsilon_1(p_3, \lambda_1)$ and
$\varepsilon_2(p_4, \lambda_2)$, have been suppressed.

Even though the representations in Eq.~($15$) have been used internally,
we present our result only for the amplitude squared and summed over 
helicities.

\section{Infrared Pole Structure}
\label{sec:catani}

In the case of one-loop  QCD amplitudes,
their poles in $\ep$ can be  
expressed as a universal combination of the tree amplitude 
and a colour-charge operator $\Ione{}$. 
The generic form of $\Ione{}$ was found by Catani and 
Seymour~\cite{Catani:1996vz} 
and it was derived for the general one-loop QCD 
amplitude by integrating the real radiation graphs of the same order  in
perturbation series in the one-particle unresolved limit.

The pole structure of our one-loop expression is given, 
according to the prediction by Catani, by acting with the operator
$\Ione{}$ onto the tree-level result:
\bea
| {\cal M}^{(1)} \rangle &=& \Ione{}  | {\cal M}^{(0)} \rangle + | {\cal
  M}^{(1)}_{\rm finite} \rangle \, ,
\eea
where $\Ione{}$ is defined as
\begin{eqnarray}
\Ione{} = -\CF \frac{e^{\ep \gamma}}{\Gamma(1-\ep)} 
\left(\frac{1}{\ep^2} + \frac{3}{2\ep} \right) 
\left(-\frac{\mu^2}{s}\right)^\ep\, .
\label{eq:Ioneqq}
\end{eqnarray}
The expression for the one-loop squared result then takes the form:
\bea
\label{eq:fini}
\hspace{-.5cm}
\A^{\nnlo \, (1 \times 1)} &=&
\langle {\cal M}^{(1)} | {\cal M}^{(1)} \rangle  \\ \nonumber
&=&  |\Ione{}|^2 \langle {\cal
  M}^{(0)} | {\cal M}^{(0)} \rangle  
+ 2  {\rm Re} \left [ \Ione{}^{*}  \langle {\cal
  M}^{(0)} | {\cal M}^{(1)}_{\rm finite} \rangle \right ]  +  \langle {\cal
  M}^{(1)}_{\rm finite} | {\cal M}^{(1)}_{\rm finite} \rangle \, .
\eea

We checked that our result is in agreement with the Catani prediction.

\section{Results}
\label{sec:results}

In this section, we give explicit expressions for the finite remainder of 
the one-loop squared contribution ${\cal F}_{\!inite,}^{( 1 \times 1)}$
defined as 
\begin{equation}
 {\cal F}_{\!inite,}^{( 1 \times 1)}(s, t, u, m, \mu) = \A^{\nnlo \, 
(1 \times 1)}(s, t, u, m, \mu) 
     - \mathcal{C}_{atani}^{(1 \times 1)}(s, t, u, m, \mu)\, ,
\end{equation} 
where $\mathcal{C}_{atani}^{(1 \times 1)}(s, t, u, m, \mu)$
is given by the first two terms of Eq.~(\ref{eq:fini}) and 
is expanded through to $\mathcal{O}(1)$, which means that
it contains finite contributions as well.

The EW structure of the finite remainder for a down-type quark
can be factorised as 
\begin{eqnarray}
\label{eq:down}
{\cal F}_{\!inite, \,\, down}^{\, (1 \times 1)} &=& 
N \, \cf^2
\sum_{i=1, 3}\, c^{}_{i}  
{\cal J}_{i,\,down}^{(1 \times 1)}(m_s, x, \frac{s}{\mu^2})  \, .
\end{eqnarray}
This decomposition allows one 
to easily obtain the result for the up-type quark
scattering. The latter is then given by
\begin{eqnarray}
\label{eq:up}
{\cal F}_{\!inite, \,\, up}^{\, (1 \times 1)} &=& 
N \, \cf^2
\sum_{i=1, 3}\, c^{}_{i}  
{\cal J}_{i,\,up}^{(1 \times 1)}(m_s, x, \frac{s}{\mu^2})  \, ,
\end{eqnarray}
where one needs to use the following formulae ($y\,=\,-\frac{u}{s}$)
\begin{equation}
{\cal J}_{1,\,up}^{(1 \times 1)}(m_s, x, \frac{s}{\mu^2})\,=
\,{\cal J}_{1,\,down}^{(1 \times 1)}(m_s, y, \frac{s}{\mu^2})\, ,
\end{equation}
\begin{equation}
{\cal J}_{2,\,up}^{(1 \times 1)}(m_s, x, \frac{s}{\mu^2})\,=
\,- {\cal J}_{2,\,down}^{(1 \times 1)}(m_s, y, \frac{s}{\mu^2})\, ,
\end{equation}
\begin{equation}
{\cal J}_{3,\,up}^{(1 \times 1)}(m_s, x, \frac{s}{\mu^2})\,=
\,{\cal J}_{3,\,down}^{(1 \times 1)}(m_s, y, \frac{s}{\mu^2})\, , 
\end{equation}
and naturally to make the corresponding changes in the definitions of
the couplings $c_1$, $c_2$ and $c_3$ namely to 
use the up-type quark charge
and isospin.  In the following we will
suppress all indices that indicate the type of scattered quark.

Our result reads:
\bea
{\mathcal J_1}^{(1 \times 1)} &=&
\frac{64 (1-x) x}{\text{m_s}^2}
+
\frac{192}{\text{m_s}}
+
\left[
-4 \left(-\frac{2}{x}-\frac{1}{x^2}-\frac{1}{x^3}+1-\frac{1}{1-x}\right)
   \ly^4+8 \left(\frac{3}{x}+\frac{2}{x^2}+\frac{3}{1-x}\right)
   \ly^3
\brk
+\left(4 \left(\frac{6}{x}+15-\frac{5}{1-x}\right)-16
   \left(-\frac{2}{x}-\frac{1}{x^2}-\frac{1}{x^3}+1-\frac{1}{1-x}\right)
   \pi^2\right) \ly^2
\brk
+\left(16
   \left(\frac{3}{x}+\frac{2}{x^2}+\frac{3}{1-x}\right) \pi^2+8
   \left(7-\frac{5}{1-x}\right)\right) \ly-4
   \left(-\frac{4}{x}+1-\frac{9}{1-x}\right) \pi^2
\brk
-4 \left(-32
   x-\frac{83}{1-x}+83\right)+128 \left(x-\frac{1}{1-x}+2\right)
   \lm
\right]
\, ,\nonumber \\
{\mathcal J_2}^{(1 \times 1)} &=& 
\frac{64 (1-x) x}{\text{m_s}^2}
+
\frac{1}{\text{m_s}}
\left[
64 \left(2 x^2-2 x+3\right)
\right]
+
\left[
-32 (x-2) \ly^2-32 \left(x-\frac{2}{1-x}+2\right)
   \ly
\brk
-32 \left(-9 x-\frac{2}{1-x}+14\right)+64
   \left(x-\frac{1}{1-x}+2\right) \lm
\right]
\, , \nonumber \\
{\mathcal J_3}^{(1 \times 1)} &=&
\frac{32 (1-x) x}{\text{m_s}^2}
+
\frac{1}{\text{m_s}}
\left[
32 \left(4 x^2-4 x+3\right)
\right]
+
\left[
-384 \left(x^2-x+1\right)
\right] \, ,
\eea
where ${\rm L_m}$ and ${\rm L_y}$ are defined as
\bea
\label{eq:lmly}
{\rm L_m} = {\rm Log}\left( m_s \right )\, , 
\quad {\rm L_y} = {\rm Log} \left ( 1-x\right ) \, .
\eea

\section{Conclusions}
\label{sec:conclusions}

We have computed the one-loop squared 
${\mathcal O} (\alpha_s^2)$ corrections
to the process $q{\bar q} \rightarrow W^+ W^-$ in a high energy expansion 
through to the
zeroth-order in $\frac{M_W^2}{s}$. 
We checked that the infrared structure of our result
agrees with the prediction of Catani's
formalism for the  infrared structure of QCD amplitudes.

The present result, given as the finite remainder of the NNLO 
one-loop squared virtual corrections after 
subtraction of the structure predicted 
by Catani's formalism, in combination with 
the result in Ref~\cite{Chachamis:2008yb} for the two-loop amplitude,
completes the calculation of the virtual
corrections to the process in the high energy limit. 
In a forthcoming publication, we will derive a series
expansion in the mass and integrate both results numerically
similarly to what has been done for top quark pair 
production~\cite{Czakon:2008zk}.

To complete the NNLO project one still needs to consider 
$2 \to 3$ real-virtual contributions  
and $2 \to 4$ real ones. The real-virtual corrections
are known from the NLO studies on $W W + jet$ production
in Refs.~\cite{Campbell:2007ev,
Dittmaier:2007th}.

{\bf{Acknowledgments:}}

    This work was supported by the Sofja Kovalevskaja Award of the
    Alexander von Humboldt Foundation and by the German Federal
    Ministry of Education and Research (BMBF) under contract number 05HT1WWA2.
%

\setcounter{section}{0}
\renewcommand\thesection{Appendix:}

\section{$\langle {\cal
  M}^{(0)} | {\cal M}^{(1)}_{\rm finite} \rangle $ to order $\epsilon^2$}
\label{app:oneloopep2}

Here we present the expression for the one-loop result,
$\langle {\cal  M}^{(0)} | {\cal M}^{(1)}_{\rm finite} \rangle$ 
up to order $\epsilon^2$
for down-type quarks. 
This result completes the list of the elements needed
in Eq.~(\ref{eq:Aexp}) in order to have the perturbative expansion of the 
amplitude up to order $\alpha_s^2$ in the high energy limit.

\bea
\langle {\cal  M}^{(0)} | {\cal M}^{(1)}_{\rm
  finite} \rangle &=& N C_F \sum_{i= 1,3} c_i {\mathcal J^{(0 \times 1)}_i} \, .
\eea

{\scalefont{0.9}

\bea
{\mathcal J^{(0 \times 1)}_1} &=& 
\left \{
\frac{1}{m_s^2}
\left[
-16 (1-x) x
\right]
+
\frac{1}{m_s}
\left[
-48
\right]
+
\left[
-8 \left(1-\frac{1}{1-x}\right) \ly^2-8
   \left(1-\frac{1}{1-x}\right) \ly
\ibrk
+8 \left(-2
   x-\frac{9}{1-x}+9\right)-16 \left(x-\frac{1}{1-x}+2\right)
   \lm
\right]
\right \} \nonumber \\ &&
+ i \pi \left \{
-16 \ly \left(1-\frac{1}{1-x}\right)-8
   \left(1-\frac{1}{1-x}\right)
\right \} \nonumber \\ &&
+\epsilon 
\left \{
\frac{1}{m_s^2}
\left[
-32 (1-x) x+8 (1-x) \z3 x+16 (1-x) \ls x
\right]
+
\frac{1}{m_s}
\left[
24 \z3+48 \ls-32
\right]
\brk
+
\left[
\frac{16}{3} \left(1-\frac{1}{1-x}\right) \ly^3+4
   \left(3-\frac{5}{1-x}\right) \ly^2+8
   \left(1-\frac{1}{1-x}\right) \ls \ly^2+
\ibrk
\left(8
   \left(1-\frac{1}{1-x}\right) \pi^2
-8
   \left(1+\frac{1}{1-x}\right)\right) \ly+8
   \left(1-\frac{1}{1-x}\right) \ls \ly+4
   \left(1-\frac{1}{1-x}\right) \pi^2
\ibrk
+8
   \left(x-\frac{1}{1-x}+2\right) \lm^2-32
   \left(x+\left(1-\frac{1}{1-x}\right) \z3\right)-16
   \left(x-\frac{2}{1-x}+4\right) \lm
\ibrk
-8 \left(-2
   x-\frac{9}{1-x}+9\right) \ls
+16
   \left(x-\frac{1}{1-x}+2\right) \lm \ls-16
   \left(1-\frac{1}{1-x}\right) \s12
\right]
\right \} \nonumber \\ &&
+ \epsilon \,i \pi \left \{
\frac{1}{m_s^2}
\left[
-16 (1-x) x
\right]
+
\frac{1}{m_s}
\left[
-48
\right]
+
\left[
8 \left(1-\frac{1}{1-x}\right) \ly^2+16
   \left(1-\frac{2}{1-x}\right) \ly
\ibrk
+16
   \left(1-\frac{1}{1-x}\right) \ls \ly
+16
   \left(-x-\frac{5}{1-x}+4\right)+16 \left(1-\frac{1}{1-x}\right)
   \li2
\ibrk
-16 \left(x-\frac{1}{1-x}+2\right) \lm+8
   \left(1-\frac{1}{1-x}\right) \ls
\right]
\right \} \nonumber \\ &&
+ \epsilon^2 
\left \{
\frac{1}{m_s^2}
\left[
\frac{2}{15} (1-x) x \pi^4+\frac{28}{3} (1-x) x
   \pi^2-8 (1-x) x \ls^2-64 (1-x) x+32 (1-x) x
   \ls
\ibrk
+\z3 (12 (1-x) x-8 (1-x) x \ls)
\right]
+
\frac{1}{m_s}
\left[
\frac{2 \pi^4}{5}+28 \pi^2-24
   \ls^2+\z3 (4-24 \ls)+32
   \ls-64
\right]
\brk
+
\left[
-\frac{8}{15} \left(1-\frac{1}{1-x}\right) \pi^4-\frac{2}{3}
   \left(-14 x-\frac{69}{1-x}+57\right) \pi^2-8
   \left(1-\frac{1}{1-x}\right) \li2 \pi^2
\ibrk
-2
   \left(1-\frac{1}{1-x}\right) \ly^4-\frac{8}{3}
   \left(x-\frac{1}{1-x}+2\right) \lm^3-\frac{4}{3}
   \left(5-\frac{9}{1-x}\right) \ly^3-\frac{16}{3}
   \left(1-\frac{1}{1-x}\right) \ls \ly^3
\ibrk
+8
   \left(x-\frac{2}{1-x}+4\right) \lm^2+4 \left(-2
   x-\frac{9}{1-x}+9\right) \ls^2-8
   \left(x-\frac{1}{1-x}+2\right) \lm \ls^2
\ibrk
-4
   \left(1-\frac{1}{1-x}\right) \ls^2
   \ly^2+\left(12 \left(1+\frac{1}{1-x}\right)-\frac{10}{3}
   \left(1-\frac{1}{1-x}\right) \pi^2\right)
   \ly^2-4 \left(3-\frac{5}{1-x}\right) \ls
   \ly^2
\ibrk
+8 \left(-8 x+\left(2-\frac{2}{1-x}\right)
   \z3-\frac{9}{1-x}+9\right)+\left(\frac{28}{3}
   \left(x-\frac{1}{1-x}+2\right) \pi^2-16 \left(2
   x-\frac{3}{1-x}+6\right)\right) \lm
\ibrk
-8
   \left(x-\frac{1}{1-x}+2\right) \lm^2
   \ls+\left(-4 \left(1-\frac{1}{1-x}\right)
   \pi^2+32 x+\left(32-\frac{32}{1-x}\right) \z3\right)
   \ls
\ibrk
+16 \left(x-\frac{2}{1-x}+4\right) \lm
   \ls-4 \left(1-\frac{1}{1-x}\right) \ls^2
   \ly+\left(-\frac{2}{3} \left(11-\frac{23}{1-x}\right)
   \pi^2-24 \left(1+\frac{1}{1-x}\right)\right)
   \ly
\ibrk
+\left(8 \left(1+\frac{1}{1-x}\right)-8
   \left(1-\frac{1}{1-x}\right) \pi^2\right) \ls
   \ly+16 \left(1-\frac{2}{1-x}\right) \s12+16
   \left(1-\frac{1}{1-x}\right) \ls \s12
\ibrk
+16
   \left(1-\frac{1}{1-x}\right) \ly \s12+16
   \left(1-\frac{1}{1-x}\right) \s13-16
   \left(1-\frac{1}{1-x}\right) \s22
\right]
\right \} \nonumber \\ &&
+ \epsilon^2 \, i \pi
\left \{
\frac{1}{m_s^2}
\left[
-32 (1-x) x+8 (1-x) \z3 x+16 (1-x) \ls x
\right]
+
\frac{1}{m_s}
\left[
24 \z3+48 \ls-32
\right]
\brk
+
\left[
-\frac{8}{3} \left(1-\frac{1}{1-x}\right) \ly^3-8
   \left(1-\frac{2}{1-x}\right) \ly^2-8
   \left(1-\frac{1}{1-x}\right) \ls \ly^2-8
   \left(1-\frac{1}{1-x}\right) \ls^2 \ly
\ibrk
+\left(4
   \left(1-\frac{1}{1-x}\right) \pi^2+16
   \left(1+\frac{1}{1-x}\right)\right) \ly-16
   \left(1-\frac{1}{1-x}\right) \li2 \ly-16
   \left(1-\frac{2}{1-x}\right) \ls \ly
\ibrk
+2
   \left(1-\frac{1}{1-x}\right) \pi^2+8
   \left(x-\frac{1}{1-x}+2\right) \lm^2-4
   \left(1-\frac{1}{1-x}\right) \ls^2
\ibrk
-8 \left(4
   x+\left(4-\frac{4}{1-x}\right) \z3+\frac{3}{1-x}+3\right)-16
   \left(1-\frac{2}{1-x}\right) \li2+16
   \left(1-\frac{1}{1-x}\right) \li3
\ibrk
-16
   \left(x-\frac{2}{1-x}+4\right) \lm-16
   \left(-x-\frac{5}{1-x}+4\right) \ls-16
   \left(1-\frac{1}{1-x}\right) \li2 \ls
\ibrk
+16
   \left(x-\frac{1}{1-x}+2\right) \lm \ls-16
   \left(1-\frac{1}{1-x}\right) \s12
\right]
\right \} \, ,
\eea

\bea
{\mathcal J^{(0 \times 1)}_2} &=& 
\left \{ 
\frac{1}{m_s^2}
\left[
-16 (1-x) x
\right]
+
\frac{1}{m_s}
\left[
-16 \left(2 x^2-2 x+3\right)
\right]
+
\left[
4 (x-2) \ly^2+4 \left(x-\frac{2}{1-x}+2\right)
   \ly
\ibrk
+4 \left(-17 x-\frac{2}{1-x}+26\right)-8
   \left(x-\frac{1}{1-x}+2\right) \lm
\right]
\right \} \nonumber \\ && 
+ i \pi
\left \{
\left[
4 \left(x-\frac{2}{1-x}+2\right)+8 (x-2) \ly
\right]
\right \} \nonumber \\ &&
+\epsilon 
\left \{
\frac{1}{m_s^2}
\left[
-32 (1-x) x+8 (1-x) \z3 x+16 (1-x) \ls x
\right]
+
\frac{1}{m_s}
\left[
-32 \left(2 x^2-2 x+1\right)
\ibrk
+8 \left(2 x^2-2 x+3\right) \z3+16 \left(2
   x^2-2 x+3\right) \ls
\right]
+
\left[
-\frac{8}{3} (x-2) \ly^3+2 \left(\frac{2}{1-x}-5 x\right)
   \ly^2
\ibrk
-4 (x-2) \ls \ly^2
+\left(-4
   (x-2) \pi^2-8 \left(\frac{1}{1-x}-2 x\right)\right)
   \ly-4 \left(x-\frac{2}{1-x}+2\right) \ls
   \ly
\ibrk
-2 \left(x-\frac{2}{1-x}+2\right) \pi^2+4
   \left(x-\frac{1}{1-x}+2\right) \lm^2+4 \left(-17 x+(8 x-12)
   \z3-\frac{2}{1-x}+18\right)
\ibrk
-8 \left(x-\frac{2}{1-x}+4\right)
   \lm-4 \left(-17 x-\frac{2}{1-x}+26\right) \ls+8
   \left(x-\frac{1}{1-x}+2\right) \lm \ls+8 (x-2)
   \s12
\right]
 \right \} \nonumber \\ &&
+ \epsilon \, i \pi
\left \{
\frac{1}{m_s^2}
\left[
-16 (1-x) x
\right]
+
\frac{1}{m_s}
\left[
-16 \left(2 x^2-2 x+3\right)
\right]
+
\left[
-4 (x-2) \ly^2-8 (2 x-1) \ly
\ibrk
-8 (x-2)
   \ls \ly+4 \left(-13 x-\frac{4}{1-x}+26\right)-8
   (x-2) \li2-8 \left(x-\frac{1}{1-x}+2\right)
   \lm
\ibrk
-4 \left(x-\frac{2}{1-x}+2\right) \ls
\right]
\right \} \nonumber \\ &&
+ \epsilon^2 
\left \{ 
\frac{1}{m_s^2}
\left[
\frac{2}{15} (1-x) x \pi^4+\frac{28}{3} (1-x) x
   \pi^2-8 (1-x) x \ls^2-64 (1-x) x+32 (1-x) x
   \ls
\ibrk
+\z3 (12 (1-x) x-8 (1-x) x \ls)
\right]
+
\frac{1}{m_s}
\left[
\frac{2}{15} \left(2 x^2-2 x+3\right) \pi^4+\frac{28}{3}
   \left(2 x^2-2 x+3\right) \pi^2
\ibrk
-8 \left(2 x^2-2 x+3\right)
   \ls^2-64 \left(2 x^2-2 x+1\right)+32 \left(2 x^2-2 x+1\right)
   \ls+\z3 \left(4 \left(6 x^2-6 x+1\right)
\dibrk
-8 \left(2
   x^2-2 x+3\right) \ls\right)
\right]
+
\left[
\frac{4}{15} (2 x-3) \pi^4+\frac{1}{3} \left(95
   x+\frac{26}{1-x}-182\right) \pi^2
\ibrk
+4 (x-2) \li2
   \pi^2+(x-2) \ly^4-\frac{4}{3}
   \left(x-\frac{1}{1-x}+2\right) \lm^3-\frac{2}{3} \left(-9
   x+\frac{2}{1-x}+2\right) \ly^3
\ibrk
+\frac{8}{3} (x-2)
   \ls \ly^3+4 \left(x-\frac{2}{1-x}+4\right)
   \lm^2+2 \left(-17 x-\frac{2}{1-x}+26\right)
   \ls^2
\ibrk
-4 \left(x-\frac{1}{1-x}+2\right) \lm
   \ls^2+2 (x-2) \ls^2
   \ly^2
+\left(\frac{5}{3} (x-2) \pi^2-4
   \left(x-\frac{1}{1-x}+1\right)\right) \ly^2
\ibrk
-2
   \left(\frac{2}{1-x}-5 x\right) \ls \ly^2+8
   \left(-17 x+(2 x-1) \z3-\frac{2}{1-x}+18\right)+\left(\frac{14}{3}
   \left(x-\frac{1}{1-x}+2\right) \pi^2
\dibrk
-8 \left(2
   x-\frac{3}{1-x}+6\right)\right) \lm-4
   \left(x-\frac{1}{1-x}+2\right) \lm^2
   \ls+\left(2 \left(x-\frac{2}{1-x}+2\right)
   \pi^2+68 x
\dibrk
+(48-32 x) \z3+\frac{8}{1-x}-72\right)
   \ls+8 \left(x-\frac{2}{1-x}+4\right) \lm
   \ls+2 \left(x-\frac{2}{1-x}+2\right) \ls^2
   \ly
\ibrk
+\left(\frac{1}{3} \left(23 x+\frac{2}{1-x}-14\right)
   \pi^2-4 \left(-5 x+\frac{4}{1-x}+2\right)\right)
   \ly
\ibrk
+\left(4 (x-2) \pi^2+8
   \left(\frac{1}{1-x}-2 x\right)\right) \ls \ly-8
   (2 x-1) \s12-8 (x-2) \ls \s12
\ibrk
-8
   (x-2) \ly \s12-8 (x-2) \s13+8
   (x-2) \s22
\right]
\right \} \nonumber \\ &&
+ \epsilon^2 \, i \pi
\left \{
\frac{1}{m_s^2}
\left[
-32 (1-x) x+8 (1-x) \z3 x+16 (1-x) \ls x
\right]
+
\frac{1}{m_s}
\left[
-32 \left(2 x^2-2 x+1\right)
\ibrk
+8 \left(2 x^2-2 x+3\right) \z3+16 \left(2
   x^2-2 x+3\right) \ls
\right]
+
\left[
\frac{4}{3} (x-2) \ly^3+4 (2 x-1) \ly^2+4 (x-2)
   \ls \ly^2
\ibrk
+4 (x-2) \ls^2
   \ly+\left(8 (x-1)-2 (x-2) \pi^2\right)
   \ly+8 (x-2) \li2 \ly+8 (2 x-1)
   \ls \ly
\ibrk
+\left(-x+\frac{2}{1-x}-2\right)
   \pi^2+4 \left(x-\frac{1}{1-x}+2\right) \lm^2+2
   \left(x-\frac{2}{1-x}+2\right) \ls^2
\ibrk
+8 \left(-6 x+(4 x-6)
   \z3-\frac{3}{1-x}+8\right)+8 (2 x-1) \li2-8 (x-2)
   \li3
\ibrk
-8 \left(x-\frac{2}{1-x}+4\right) \lm-4
   \left(-13 x-\frac{4}{1-x}+26\right) \ls+8 (x-2)
   \li2 \ls
\ibrk
+8 \left(x-\frac{1}{1-x}+2\right)
   \lm \ls+8 (x-2) \s12
\right]
\right \} \, ,
\eea

\bea
{\mathcal J^{(0 \times 1)}_3} &=& 
\left \{
\frac{1}{m_s^2}
\left[
-8 (1-x) x
\right]
+
\frac{1}{m_s}
\left[
-8 \left(4 x^2-4 x+3\right)
\right]
+
\left[
96 \left(x^2-x+1\right)
\right]
\right \} \nonumber \\ &&
+\epsilon 
\left \{
\frac{1}{m_s^2}
\left[
-16 (1-x) x+4 (1-x) \z3 x+8 (1-x) \ls x
\right]
+
\frac{1}{m_s}
\left[
-16 \left(4 x^2-4 x+1\right)
\ibrk
+4 \left(4 x^2-4 x+3\right) \z3+8 \left(4
   x^2-4 x+3\right) \ls
\right]
+
\left[
-16 \left(-8 x^2+8 x+\left(3 x^2-3 x+3\right) \z3-4\right)
\ibrk
-96
   \left(x^2-x+1\right) \ls
\right]
\right \} \nonumber \\ &&
+\epsilon \, i \pi
\left \{
\frac{1}{m_s^2}
\left[
-8 (1-x) x
\right]
+
\frac{1}{m_s}
\left[
-8 \left(4 x^2-4 x+3\right)
\right]
+
\left[
96 \left(x^2-x+1\right)
\right]
\right \} \nonumber \\ &&
+ \epsilon^2 
\left \{ 
\frac{1}{m_s^2}
\left[
\frac{1}{15} (1-x) x \pi^4+\frac{14}{3} (1-x) x
   \pi^2-4 (1-x) x \ls^2-32 (1-x) x+16 (1-x) x
   \ls
\ibrk
+\z3 (6 (1-x) x-4 (1-x) x \ls)
\right]
+
\frac{1}{m_s}
\left[
\frac{1}{15} \left(4 x^2-4 x+3\right) \pi^4+\frac{14}{3}
   \left(4 x^2-4 x+3\right) \pi^2
\ibrk
-4 \left(4 x^2-4 x+3\right)
   \ls^2-32 \left(4 x^2-4 x+1\right)+16 \left(4 x^2-4 x+1\right)
   \ls+\z3 \left(2 \left(12 x^2-12 x+1\right)
\dibrk
-4 \left(4
   x^2-4 x+3\right) \ls\right)
\right]
+
\left[
-\frac{4}{5} \left(x^2-x+1\right) \pi^4-56 \left(x^2-x+1\right)
   \pi^2+48 \left(x^2-x+1\right) \ls^2
\ibrk
-8
   \left(-32 x^2+32 x+\left(5 x^2-5 x+1\right) \z3-16\right)+16 \left(-8
   x^2+8 x+\left(3 x^2-3 x+3\right) \z3-4\right) \ls
\right]
\right \} \nonumber \\ &&
+ \epsilon^2 \, i \pi 
\left \{ 
\frac{1}{m_s^2}
\left[
-16 (1-x) x+4 (1-x) \z3 x+8 (1-x) \ls x
\right]
+
\frac{1}{m_s}
\left[
-16 \left(4 x^2-4 x+1\right)
\ibrk
+4 \left(4 x^2-4 x+3\right) \z3+8 \left(4
   x^2-4 x+3\right) \ls
\right]
+
\left[
-16 \left(-8 x^2+8 x+\left(3 x^2-3 x+3\right) \z3-4\right)
\ibrk
-96
   \left(x^2-x+1\right) \ls
\right]
\right \} \, ,
\eea

}
where ${\rm L_m}$ and ${\rm L_y}$ are defined in Eq.~(\ref{eq:lmly})
and
\bea
{\rm L_s} = {\rm Log} \left ( \frac{s}{\mu^2}
\right ) \, .
\eea


{\footnotesize



\begin{thebibliography}{10}

\bibitem{tevatron}
CDF Collaboration, Phys. Rev. Lett. {\bf 94} (2005) 211801;
\newblock D0 Collaboration, Phys. Rev. Lett. 94 (2005) 151801

\bibitem{Spira:1995rr}
M.~Spira, A.~Djouadi, D.~Graudenz and P.~M.~Zerwas,
\newblock  Nucl.\ Phys.\  B {\bf 453} (1995) 17

\bibitem{Dawson:1990zj}
S.~Dawson,
\newblock  Nucl.\ Phys.\  B {\bf 359} (1991) 283.

\bibitem{Harlander:2002wh}
R.~V.~Harlander and W.~B.~Kilgore,
\newblock  Phys.\ Rev.\ Lett.\  {\bf 88} (2002) 201801

\bibitem{Anastasiou:2002yz}
C.~Anastasiou and K.~Melnikov,
\newblock  Nucl.\ Phys.\  B {\bf 646} (2002) 220

\bibitem{Ravindran:2003um}
V.~Ravindran, J.~Smith and W.~L.~van Neerven,
\newblock  Nucl.\ Phys.\  B {\bf 665} (2003) 325

\bibitem{Catani:2001cr}
S.~Catani, D.~de Florian and M.~Grazzini,
\newblock  JHEP {\bf 0201} (2002) 015

\bibitem{Davatz:2004zg}
G.~Davatz, G.~Dissertori, M.~Dittmar, M.~Grazzini and F.~Pauss,
\newblock  JHEP {\bf 0405} (2004) 009

\bibitem{Anastasiou:2004xq}
C.~Anastasiou, K.~Melnikov and F.~Petriello,
\newblock  Phys.\ Rev.\ Lett.\  {\bf 93} (2004) 262002

\bibitem{Anastasiou:2007mz}
C.~Anastasiou, G.~Dissertori and F.~Stockli,
\newblock  arXiv:0707.2373 [hep-ph].
\newblock  

\bibitem{Grazzini:2008tf}
M.~Grazzini,
\newblock arXiv: 0801.3232 [hep-ph]


\bibitem{Bredenstein:2006rh}
A.~Bredenstein, A.~Denner, S.~Dittmaier and M.~M.~Weber,
\newblock  Phys.\ Rev.\  D {\bf 74} (2006) 013004

\bibitem{kauer1}
T.~Binoth, M.~Ciccolini, N.~Kauer and M.~Kramer,
\newblock  JHEP {\bf 0503} (2005) 065  [arXiv:hep-ph/0503094].
\newblock  

\bibitem{kauer2}
T.~Binoth, M.~Ciccolini, N.~Kauer and M.~Kramer,
\newblock  JHEP {\bf 0612} (2006) 046  [arXiv:hep-ph/0611170].
\newblock  

\bibitem{dittmardreiner}
M.~Dittmar and H.~K.~Dreiner,
\newblock  Phys.\ Rev.\  D {\bf 55} (1997) 167 [arXiv:hep-ph/9608317].
\newblock  

\bibitem{brown}
R.~W.~Brown and K.~O.~Mikaelian,
\newblock  Phys.\ Rev.\  D {\bf 19} (1979) 922.
\newblock  

\bibitem{ohn}
J.~Ohnemus,
\newblock  Phys.\ Rev.\  D {\bf 44} (1991) 1403.
\newblock  

\bibitem{fri}
S.~Frixione,
\newblock  Nucl.\ Phys.\  B {\bf 410} (1993) 280.
\newblock  

\bibitem{dixon1}
L.~J.~Dixon, Z.~Kunszt and A.~Signer,
\newblock  Nucl.\ Phys.\  B {\bf 531} (1998) 3  [arXiv:hep-ph/9803250].
\newblock  

\bibitem{dixon2}
L.~J.~Dixon, Z.~Kunszt and A.~Signer,
\newblock  Phys.\ Rev.\  D {\bf 60} (1999) 114037  [arXiv:hep-ph/9907305].
\newblock  

\bibitem{campbell}
J.~M.~Campbell and R.~K.~Ellis,
\newblock  Phys.\ Rev.\  D {\bf 60} (1999) 113006  [arXiv:hep-ph/9905386].
\newblock  

\bibitem{grazzini:2006}
M. Grazzini,
\newblock JHEP {\bf 0601} (2006) 095
\newblock 

\bibitem{Anastasiou:2002zn}
C.~Anastasiou, E.~W.~N.~Glover and M.~E.~Tejeda-Yeomans,
\newblock  Nucl.\ Phys.\  B {\bf 629} (2002) 255


\bibitem{Chachamis:2008yb}
G.~Chachamis, M.~Czakon and D.~Eiras,
\newblock arXiv:0802.4028 [hep-ph]

\bibitem{Chachamis:2007cy}
  G.~Chachamis,
  Acta Phys.\ Polon.\  B {\bf 38} (2007) 3563
  [arXiv:0710.3035 [hep-ph]].


\bibitem{qqTT}
M. Czakon, A. Mitov and S. Moch,
\newblock Phys. Lett. B651 (2007) 147, arXiv:0705.1975 [hep-ph]
\newblock 

\bibitem{ggTT}
M.~Czakon, A.~Mitov and S.~Moch,
\newblock   arXiv:0707.4139 [hep-ph].
\newblock   

\bibitem{Czakon:2004wm}
M. Czakon, J. Gluza and T. Riemann,
\newblock Phys. Rev. D71 (2005) 073009, hep-ph/0412164
\newblock 

\bibitem{Czakon:2006pa}
M. Czakon, J. Gluza and T. Riemann,
\newblock Nucl. Phys. B751 (2006) 1, hep-ph/0604101
\newblock 

\bibitem{Actis:2007gi}
S.~Actis, M.~Czakon, J.~Gluza and T.~Riemann,
\newblock Nucl.\ Phys.\  B {\bf 786} (2007) 26


\bibitem{Smirnov:1999gc}
V.A. Smirnov,
\newblock Phys. Lett. B460 (1999) 397, hep-ph/9905323
\newblock 


\bibitem{Tausk:1999vh}
J.B. Tausk,
\newblock Phys. Lett. B469 (1999) 225, hep-ph/9909506
\newblock 


\bibitem{MBrepresentation}
G. Chachamis and M. Czakon,
\newblock {\tt MBrepresentation.m}, Unpublished


\bibitem{Czakon:2005rk}
M. Czakon,
\newblock Comput. Phys. Commun. 175 (2006) 559, hep-ph/0511200
\newblock 


\bibitem{Moch:2005uc}
S. Moch and P. Uwer,
\newblock Comput. Phys. Commun. 174 (2006) 759, math-ph/0508008
\newblock 


\bibitem{pslq:1992}
H.R.P. Ferguson and D.H. Bailey,
\newblock (1992), (see e.g. http://mathworld.wolfram.com/PSLQAlgorithm.html).

\bibitem{catani}
S.~Catani,
\newblock  Phys.\ Lett.\  B {\bf 427} (1998) 161  [arXiv:hep-ph/9802439].
\newblock


\bibitem{vanRitbergen:1997va}
T. van Ritbergen, J.A.M. Vermaseren and S.A. Larin,
\newblock Phys.\ Lett.\ B {\bf 400} (1997) 379
\newblock 


\bibitem{Czakon:2004bu}
M. Czakon,
\newblock Nucl. Phys. B710 (2005) 485, hep-ph/0411261
\newblock 


\bibitem{Diener:1997nx}
K.~P.~O.~Diener, B.~A.~Kniehl and A.~Pilaftsis,
\newblock Phys.\ Rev.\  D {\bf 57} (1998) 2771

\bibitem{Denner:1988tv}
A.~Denner and T.~Sack,
\newblock Nucl.\ Phys.\  B {\bf 306}, 221 (1988).


\bibitem{Catani:1996vz}
S.~Catani and M.~H.~Seymour,
\newblock  Nucl.\ Phys.\  B {\bf 485} (1997) 291
\newblock  [Erratum-ibid.\  B {\bf 510} (1998) 503]


\bibitem{Czakon:2008zk}
M.~Czakon,
\newblock  arXiv:0803.1400 [hep-ph]



\bibitem{Campbell:2007ev}
J.~M.~Campbell, R.~K.~Ellis and G.~Zanderighi,
\newblock  JHEP {\bf 0712} (2007) 056

\bibitem{Dittmaier:2007th}
S.~Dittmaier, S.~Kallweit and P.~Uwer,
\newblock Phys.\ Rev.\ Lett.\  {\bf 100} (2008) 062003




\end{thebibliography}

}
\end{document}